\documentclass[twocolumn,showpacs,showkeys,amsmath,amssymb]{revtex4}

\usepackage{graphicx}
\usepackage{amssymb}
\usepackage[latin2]{inputenc}

\def\eV{\hbox{ eV}}
\def\keV{\hbox{ keV}}
\def\MeV{\hbox{ MeV}}
\def\GeV{\hbox{ GeV}}
\def\TeV{\hbox{ TeV}}
\def\mm{\hbox{ mm}}

\def\m{\hbox{ m}}
\def\y{\hbox{ y}}

\begin{document}

\title{Majorana Neutrino, the Size of Extra Dimensions, \\ 
and Neutrinoless Double Beta Decay}
\author{Marek Góźdź} 
\email{mgozdz@kft.umcs.lublin.pl}
\author{Wiesław A. Kamiński}
\email{kaminski@neuron.umcs.lublin.pl}
\affiliation{Theoretical Physics Department, Maria Curie--Skłodowska
University, Lublin, Poland}

\begin{abstract}
The problem of Majorana neutrino mass generated in
Arkani-Hamed--Dimopoulos--Dvali model with $n$ extra spatial dimensions
is discussed. Taking into account constraints on neutrino masses coming
from cosmological observations, it is possible to obtain lower limits on
the size of extra dimensions as large as $10^{-6}\mm$. In the case of
$n=4$ it is easy to lower the fundamental scale of gravity from the
Planck energy to electroweak scale $\sim 1 \TeV$ without imposing any
additional constraints. A link between the half-life of neutrinoless
double beta decay and the size of extra dimensions is discussed.
\end{abstract}


\maketitle

\section{Introduction}

Theories and models with additional spatial dimensions have drawn much
attention during last few years (see e.g. \cite{kubyshin} and references
therein for a recent review). Such ideas have sources in the works of
T. Kaluza and O. Klein who showed in the 20's of the previous century
\cite{kaluza, klein}, that adding a fifth spatial dimension allow to
unify Maxwell's and Einstein's theories, and obtain in this way a common
description of electromagnetic and gravitational interactions. The
theory of Kaluza and Klein had been proven to be wrong and therefore the
idea of extra dimensions has been unpopular until the development of
string theory. The latter, which is now treated by many physicists as a
very promising replacement for quantum field theory and a step in the
right direction towards Theory of Everything, requires for consistent
formulation many more spatial dimensions than just three. What is more,
new multidimensional objects, called branes, emerge from this theory in
a natural way. Recent models suggest that our observable universe could
be embedded on such brane, which in turn floats in higher dimensional
bulk, possibly interacts with fields that populate the bulk as well as
with other branes. Such setup gives completely new possibilities of
solving many important issues, like the problem of hierarchy among
fundamental interactions and the smallness of neutrino mass.

At present, there are two main approaches to the problem of extra
dimensions. One of them, the Randall--Sundrum model, is based on
five-dimensional background metric solutions \cite{rs1, rs2}. This
geometric approach allows to solve the generalized Einstein
equations. Another formalism, the Arkani-Hamed--Dimopoulos--Dvali (ADD)
model (\cite{add1, add2, add3} and references therein), has more
phenomenological bases, but explains in an easy way the observed
weakness of gravitational force. It also solves the hierarchy problem by
lowering the gravitational scale from the Planck energy in $3+1$
dimensions to the electroweak scale $\sim 1 \TeV$ in $n$ additional
dimensions. It assumes further that from all known interactions only
gravity feels the extra dimensional space. The volume of extra
dimensions, in which graviton can propagate, suppresses the observed
strength of gravity. The whole Standard Model (SM) is confined to a
3-brane on which we live. Various modifications and improvements of the
model introduce other branes, which exist parallel to ours, as well as
other particles which can be found in the bulk. The possible
interactions of our brane with these objects lead to a lot of
interesting phenomena. Among others a naturally small Majorana neutrino
mass can be generated.

The first goal of the present work is to compare this mass with newest
theoretical and experimental limits, which allows to set constraints on
the sizes and number of extra dimensions. As first, we take into account
the limits on neutrino masses coming from astrophysical observations and
obtain certain values of compactification radii of extra dimensions in
ADD model. The second goal is to establish a link between the physics of
extra dimensions and theory of neutrinoless double beta decay. As is
well known, the half-life of this exotic nuclear transition heavily
depends on the masses and mixing of neutrinos, so the second source of
constraints comes from newest data from the neutrinoless double beta
decay experiments.

The paper is organized as follows. In the next section we briefly review
the ADD model and show the mechanism how the Majorana neutrino mass term
is generated. Next, in Section III, we confront the derived formula with
cosmological constraints on neutrino masses. As a result estimates of
the sizes of extra dimensions and the magnitude of the true scale of
gravity can be done. In Section IV we derive the half-life of
neutrinoless double beta decay as a function of number and size of
additional dimensions and show that the known experimental bounds on the
half-life put additional constraints on the compactification radius. A
short summary follows at the end.

\section{Majorana neutrino mass in Arkani-Hamed--Dimopoulos--Dvali model}

In ADD approach the Standard Model is localized on a three-dimensional
brane which is embedded into a $(4+n)$-dimensional space-time. The $n$
additional space-like dimensions are often referred to as being
\emph{transverse} to our brane. This construction has its origin in more
general string theoretical setting with possibly more complicated
geometry. For our purposes it is enough to assume that all the extra
dimensions are characterized by a common size $R$, so that the volume
$V_n$ of the space of extra dimensions is proportional to $R^n$. One of
the basic relations is the so-called reduction formula which can be
obtained using for example the generalized Gauss Law. It reads
\cite{add4}
\begin{equation}
  M_{Pl}^2 \sim R^n M_*^{2+n},
\end{equation}
where $M_{Pl}$ is the Planck mass and $M_*$ is the true scale of
gravity, which we want to set somewhere around 1 TeV. The coordinates
are denoted by $\{x^\mu,y^m\}$, where $\mu=0\dots 3$ labels the ordinary
space-time coordinates and $m=1 \dots n$ labels the extra
dimensions. What is more, we identify $y\sim y+2\pi R$, that is, we
compactify the extra dimensions on circles. From now on we will drop the
indices $\mu$ and $m$ for simplicity.

Assume that our brane has coordinates $\{x,0\}$ and that there is
another brane located at $\{x',y_*\}$, separated from ours by the
distance $r=|y_*|$ in the transverse dimensions. We assume further
\cite{add4, add5} that lepton number is conserved on our brane but
maximally broken on the other one. The breaking occurs in a reaction
where a particle $\chi$ with lepton number $L=2$ escapes the other brane
into the bulk. This particle, called the \emph{messenger}, may interact
with our brane and transmit to us the information about lepton number
breaking.

To be more specific, let us introduce a field $\phi_{L=2}$ located on
the other brane, whose vaccum expectation value (vev) breaks the lepton
number. What is more, it acts as a source for the bulk messenger field
$\chi$ and ``shines'' it everywhere, in particular also on our brane. We
introduce a lepton doublet $\mathcal{L}$ and a Higgs field $\mathcal{H}$
localized on our brane. They can interact with the messenger and the
interaction is given by the following action \cite{add5}:
\begin{eqnarray}
  M_*^{n-1} S^{int} \sim 
  &\int d^4x'\ \langle \phi \rangle \chi(x',y_*) + \nonumber \\
  & \int d^4x \ \alpha (lh^*)^2(x) \chi(x,0).
\label{S}
\end{eqnarray}
The first part of $S^{int}$ describes the process which takes place on
the other brane, the second one represents the interaction with SM
particles, proportional to some numerical constant $\alpha$. The
strength of the shined $\chi$ is in a natural way suppressed by the
distance $r$ between branes, and therefore one can write for the
messenger
\begin{equation}
  \langle \chi \rangle = \langle \phi \rangle \Delta_n(r),
\label{chi}
\end{equation}
where $\Delta_n(r)$ is the $n$-dimensional propagator
\begin{equation}
  \Delta_n(r) = \frac{1}{-\partial_n^2 + m_\chi^2} (r) 
  = \int d^nk \frac{e^{ikr}}{k^2 +  m_\chi^2}.
\end{equation}
The explicit form of the propagator reads 
\begin{eqnarray}
  \Delta_2(r) &\sim &
  \begin{cases}
    -\log(rm_\chi), &(rm_\chi \ll 1) \\
    \frac{e^{-rm_\chi}}{\sqrt{rm_\chi}}, &(rm_\chi \gg 1)
  \end{cases}  
  \\
  \Delta_{n>2}(r) &\sim &
  \begin{cases}
    \frac{1}{r^{n-2}}, &(rm_\chi \ll 1) \\
    \frac{e^{-rm_\chi}}{r^{n-2}}, &(rm_\chi \gg 1)
   \end{cases}
\end{eqnarray}
and one sees that it depends on the number of extra dimensions, the
distance between branes, and the mass of the messenger. This feature
will be used later. After substituting (\ref{chi}) into (\ref{S}),
writing the Higgs field in terms of its vev $v$, and identifying $l$
with $\nu_L$ we arrive at a mass term of the Majorana form
$$
m_{Maj}\ \nu_L^T \ \nu_L
$$
with the mass given approximately by
\begin{equation}
  m_{Maj} \sim \frac{v^2 \Delta(r)}{M_*^{n-1}}.
\label{mass}
\end{equation}

This is the original relation given by the authors in \cite{add5} and it
is useful if one wants to set the $M_*$ scale to a certain value, like
the electroweak scale. However, another approach is possible. Let us
place the second brane as far as possible, that is set $r=R$. Now, using
the reduction formula one can rewrite (\ref{mass}) in the form
\begin{equation}
  m_{Maj} \sim v^2 R^\frac{n(n-1)}{n+2} M_{Pl}^\frac{2(1-n)}{n+2}
  \Delta_n(R)
\label{mmaj}
\end{equation}
and deduce the size $R$ of extra dimensions imposing constraints on
neutrino mass, coming from other sources like for example cosmological
observations, neutrinoless double beta decay or neutrino oscillation
experiments.

\section{Extra dimensions from cosmological constraints}

Recently reported cosmological observations \cite{cosmo1, cosmo2} take
into account, among others, redshifts of galaxies, the microwave
background radiation, type Ia supernova behaviour and the Big Bang
nucleosynthesis, and suggest the sum of neutrino masses of all flavours
not to exceed few electronvolts
\begin{equation}
  \sum_i (m_\nu)_i \lesssim 1 \eV.
\end{equation}
Imposing this condition on (\ref{mass}) leads to a set of constraints
for the possible size of extra dimensions. Putting in numbers: $ v^2 =
(174 \GeV)^2 \sim 10^{22} \eV^2 $, $M_{Pl} \sim 10^{28} \eV$, $1
\eV^{-1} \sim 10^{-7} \m$, and using the explicit forms for the
propagator, we obtain interesting bounds on $R$.

\subsection{Two additional dimensions ($n=2$)}
For light messenger particle we obtain bounds from above. If $Rm_\chi =
0.1$, we get $R<1.8\cdot 10^{-17} \eV^{-1} \sim 10^{-24} \m$. Similarly
for $Rm_\chi = 0.01$, the bound is even stronger $R<4.7\cdot 10^{-18}
\eV^{-1} \sim 10^{-25} \m$.

If the messenger is heavy, due to exponent in the denominator of
$\Delta$ we obtain bounds from below. They turn out to be for $m_\chi =
1 \keV$, $R>1.5\cdot 10^{-2} \eV^{-1} \sim 10^{-6} \mm$.  The Newton Law
has been recently checked down to around $0.1 \mm$ \cite{long}. The
distance of $10^{-6} \mm$ is still out of range for the currently
planned tabletop experiments, but $R$ may be reached in the next
generation of projects. It shows that this case may be promising. For
heavier $\chi$ the limit goes down and takes the values for $m_\chi =
1\MeV$ $R>1.15\cdot 10^{-5} \eV^{-1} \sim 10^{-12} \m$, and for $m_\chi
= 1\GeV$ $R>8\cdot 10^{-9} \eV^{-1} \sim 10^{-16} \m$.

\subsection{Three additional dimensions ($n=3$)} 
For light messenger, performing similar analysis gives us $R<0.01 \mm$. For
heavy $\chi$ particle we can use the fact that $Rm_\chi$ must be at
least equal to one, which implies $R < 1.4 \mm$. None of these cases is
excluded and the distance 0.01 mm may be explored in not very far future. 

\subsection{Four and more additional dimensions ($n\ge 4$)}
The case of four extra dimensions and a light messenger is an
exceptional one:
$$
m_{Maj} \sim v^2 R^2 M_{Pl}^{-1} \frac{1}{R^2},
$$
because the dependence on $R$ is lost. Explicitly we get
$$
m_{Maj} \sim 10^{22} R^2 (10^{28})^{-1} \frac{1}{R^2} \eV = 10^{-6} \eV.
$$
One should stress that this result for the neutrino mass is in agreement
with all experimental facts known nowadays. An arbitrary $R$ implies
also an arbitrary value of $M_*$ which can be calculated from the
reduction formula. For example, for $M_* = 1 \TeV$ the size of extra
dimensions need to be $R \sim 10^{-8}\mm$, which is quite resonable.

The case of heavy $\chi$ does not provide any additional constraints,
namely we get $Rm_\chi > -13.8$. Similarly for more than four dimensions
we obtain $R > 10^{-99}\eV$ so all what we can say is only that these
cases are not excluded.

\section{Implications for neutrinoless double beta decay}

The presence of extra dimensions will surely influence the results of
sensitive experiments. The most obvious effect will be the violation of
inverse square law of gravity forces. In the present section we discuss
another possibility, namely the impact of extra dimensions on the
half-life of neutrinoless double beta decay.

Neutrinoless double beta decay ($0\nu 2\beta$) is a process in which a
nucleus undergoes two simultaneous beta decays without emission of
neutrinos
\begin{equation}
  A(Z,N) \to A(Z+2,N-2) + 2e^-.
\end{equation}
It requires neutrino to be a Majorana particle, so that two neutrinos
emited in beta decays annihilate with each other. It is readily seen
that this process violates the lepton number by two units, thus is
forbidden in the framework of SM. In the matter of fact, $0\nu 2\beta$
has not been observed, but the non-observability sets valuable
constraints on the shape of non-standard physics.

Ignoring the contributions from right-handed weak currents, the
half-life of $0\nu 2\beta$ can be written in the form \cite{doi}
\begin{equation}
  (T_{1/2})^{-1} = {\mathcal M} 
  \frac{\langle m_\nu \rangle^2}{m_e^2}, 
\end{equation}
where $\mathcal M$ is a nuclear matrix element which can be calculated
within some nuclear model (like the bag model or non-relativistic quark
model), and $m_e$ is the electron mass. The so-called effective neutrino
mass $\langle m_\nu \rangle$ is defined by the relation
\begin{equation}
  \langle m_\nu \rangle = \sum_i |U_{ei}^2| m_i,
\label{mnu}
\end{equation}
where $U$ is the neutrino mixing matrix and $m_i$ are neutrino mass
eigenstates. One sees that it is possible to identify $\langle m_\nu
\rangle$ with the $ee$ entry of neutrino mass matrix in the flavor basis
\begin{equation}
  \langle m_\nu \rangle = m_{ee},
\end{equation}
which is given exactly by the superposition of mass eigenstates from
equation (\ref{mnu}).

Since the mixing among neutrino mass states is not known exactly, we are
forced to simplify our picture a little bit. Using results from the
CHOOZ experiment \cite{chooz1, chooz2} we get the constraint $|U_{e3}^2|
< 0.037 - 0.017$, depending on the mixing pattern considered (two-- or
three--neutrino mixing). It implies that the contribution coming from
$m_3$ may be neglected. The remaining masses are nearly degenerate
\cite{pas-weiler} and we can approximate $m_2 \approx m_1$. Denoting the
relative phase between the two flavors by $\phi_{12}$ we obtain
\begin{equation}
  m_{ee}^2 = [1-\sin^2(2\theta_{Solar})\sin^2(\phi_{12}/2)] m_1^2.
\end{equation}
The currently favored large mixing angle (LMA) solution for solar
neutrino problem sets the value of $\sin^2(2\theta_{Solar})$ between 0.3
and 0.93 \cite{solar-nu} which may introduce a huge
uncertainty. Fortunatelly, in the final formula this uncertainty is heavily
reduced. If we assume that CP-symmetry is not violated, we
have to set $\phi_{12}$ either to 0 or $\pi/2$. The latter case is of
course more interesting and we will stick to it.

Under these conditions we are now legitimate to replace $m_1$ with
$m_{Maj}$ and link in this way the half-life of $0\nu 2\beta$ with size
of extra dimensions. We obtain
\begin{equation}
  T_{1/2}^{th} \approx (1.17 - 1.8) T_{1/2}^{exp} \frac{\langle m_\nu
  \rangle^2_{exp}}{m_{Maj}^2},
\label{t12}
\end{equation}
where $m_{Maj}$ is given by the relation (\ref{mmaj}). The
experimental values established by the IGEX collaboration \cite{igex}
are
\begin{displaymath}
  T_{1/2}^{IGEX} > 1.57 \cdot 10^{25} y, \qquad 
  \langle m_\nu \rangle_{IGEX} = (0.33 - 1.35) \eV.
\end{displaymath}
%

\begin{figure}
\includegraphics{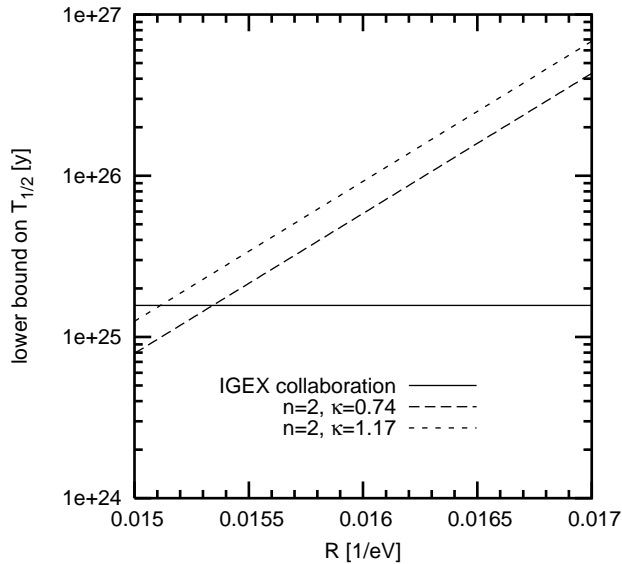}
\caption{\label{fig1} Lower bounds on the half-life of neutrinoless
  double beta decay in the case of two additional spatial dimensions of
  size $R$. The mass of the messenger particle is $m_\chi=1\keV$.}
\end{figure}

On the other hand, the Heidelberg--Moscow collaboration \cite{0nu2beta}
gave a best fit value for the effective neutrino mass $\langle m_\nu
\rangle_{H-M} = 0.39 \eV$ so in our calculations we set this parameter
to 0.4 eV. Putting in numbers and expanding the formula (\ref{t12})
using (\ref{mmaj}) we arrive at the following relation
\begin{equation}
  T_{1/2}^{th} > \kappa \cdot 10^{\frac{93n-150}{n+2}}
  R^{\frac{2n(1-n)}{n+2}} [\Delta_n(R)]^{-2} \y.
\label{t12th}
\end{equation}
Here, the uncertainty factor $\kappa$ satisfies $0.74 < \kappa < 1.17$.
For the special case of $n=2$, $m_\chi=1\keV$ formula (\ref{t12th})
simplifies to
\begin{equation}
  T_{1/2}^{th} > \kappa \cdot 10^{12} \exp(2000 R) \y,
\end{equation}
with $R > 1.5 \cdot 10^{-2} \eV^{-1}$. The result is presented on
Fig. \ref{fig1} together with bound from the experiment. One sees that
with increasing size of extra dimensions, the half-life explodes
exponentially. Therefore, the possibility of establishing $T_{1/2}$ in
experiment will set precise constraints on physics of extra
dimensions. It is visible from Fig. \ref{fig1} that for some values of
$\kappa$ the experiment pushes the size of extra dimensions even more
up: $R>0.0154 \eV^{-1}$ for $\kappa = 0.74$.

The second special case, $n=4$ with a light messenger particle, renders
the value of $T_{1/2}$ up to $\sim 10^{37} \y$ which, even if true,
remains far beyond the sensitivity of current and currently planned
experiments.

\section{Summary}

In the present paper we have shown that there are two especially
interesting cases allowed by the ADD model of extra dimensions. The
number of additional transverse dimensions $n=2$ and mass of the
messenger particle $\chi$ around $1 \keV$ implies the size of the extra
dimensions $10^{-6} \mm < R < 10^{-3} \mm$ and equivalently the true
scale of gravity $10 \TeV < M_* < 1000 \TeV$. The second case $n=4$ and
light $\chi$ suggests the mass of neutrino to be of the order of
$10^{-6} \eV$ and permits adjustments of $R$ and $M_*$ in wide range.

One of the main drawbacks of extradimensional theories is the difficulty
of experimental verification. The best known bounds are the tabletop
tests of Newton's Law. Some proposals were given to link the presence of
extra dimensions with behaviour of supenovae, however the mechanism of
supernova explosion is not known well enough to draw reasonable
conclusions. In this paper we have outlined a method how to use the
neutrinoless double beta decay for that purpose. Observation of this
exotic nuclear process will tell us precisely, which values of
parameters are allowed. Up till that time, its non-observability sets
strict constraints, which can be combined with bounds from other
sources.

\bibliography{art3}
\end{document}